*Article*

# Multi-Scale Analysis of Agricultural Drought Propagation on the Iberian Peninsula Using Non-Parametric Indices

Marco Possega [1,*], Matilde García-Valdecasas Ojeda [2,3] and Sonia Raquel Gámiz-Fortis [2,3]

1. Department of Physics and Astronomy, University of Bologna, 40126 Bologna, Italy
2. Department of Applied Physics, Faculty of Sciences, University of Granada, 18071 Granada, Spain; mgvaldecasas@ugr.es (M.G.-V.O.); srgamiz@ugr.es (S.R.G.-F.)
3. Andalusian Inter-University Institute for Earth System Research (IISTA-CEAMA), Avda. Del Mediterráneo s/n, 18006 Granada, Spain
* Correspondence: marco.possega2@unibo.it

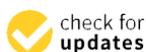

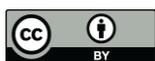

**Abstract:** Understanding how drought propagates from meteorological to agricultural drought requires further research into the combined effects of soil moisture, evapotranspiration, and precipitation, especially through the analysis of long-term data. To this end, the present study examined a multi-year reanalysis dataset (ERA5-Land) that included numerous drought events across the Iberian Peninsula, with a specific emphasis on the 2005 episode. Through this analysis, the mechanisms underlying the transition from meteorological to agricultural drought and its features for the selected region were investigated. To identify drought episodes, various non-parametric standardized drought indices were utilized. For meteorological droughts, the Standardized Precipitation-Evapotranspiration Index (SPEI) was employed, while the Standardized Soil Moisture Index (SSI), Multivariate Standardized Drought Index (MSDI), and Standard Precipitation, Evapotranspiration and Soil Moisture Index (SPESMI) were utilized for agricultural droughts, while their ability to identify relative vegetation stress in areas affected by severe droughts was investigated using the Fraction of Absorbed Photosynthetically Active Radiation (FAPAR) Anomaly provided by the Copernicus European Drought Observatory (EDO). A statistical approach based on run theory was employed to analyze several characteristics of drought propagation, such as response time scale, propagation probability, and lag time at monthly, seasonal, and six-month time scales. The retrieved response time scale was fast, about 1–2 months, and the probability of occurrence increased with the severity of the originating meteorological drought. The duration of agricultural drought was shorter than that of meteorological drought, with a delayed onset but the same term. The results obtained by multi-variate indices showed a more rapid propagation process and a tendency to identify more severe events than uni-variate indices. In general terms, agricultural indices were found to be effective in assessing vegetation stress in the Iberian Peninsula. A newly developed combined agricultural drought index was found to balance the characteristics of the other adopted indices and may be useful for future studies.

**Keywords:** drought propagation; agricultural drought; meteorological drought; Iberian Peninsula; non-parametric drought index

## 1. Introduction

Droughts are complex and spatially heterogeneous phenomena, with high variability of conditions between adjacent locations, making it easy to find an area subject to drought while neighbouring regions feature normal or even wet conditions. These spatial characteristics are mainly detectable in climatic transition areas where atmospheric influences are heterogeneous. The Iberian Peninsula (IP) is a notable example of such an area (Figure 1), a Mediterranean region located between temperate and subtropical climates, and is subject to diversified atmospheric patterns that cause a large variability of precipitation [1–3], that has presented recurrent droughts and a significant tendency towards more arid conditions in the last decades [4]. Drought is a multi-scalar phenomenon, as the effects of precipitation deficits occur across different systems and at various time scales. This is described by [5] and is due to the involvement of mechanisms at multiple scales. The drought signal is propagated through a water and energy cycle that involves a multitude of processes. The first transition in the propagation of drought generally occurs from meteorological to agricultural drought, which is driven by the response of soil moisture or crop yield to various meteorological variables such as precipitation and evapotranspiration. The combined effect of water shortage due to a lack of precipitation and the enhanced atmospheric evaporative demand can lead to a significant depletion of soil moisture, which in turn can trigger agricultural drought events.

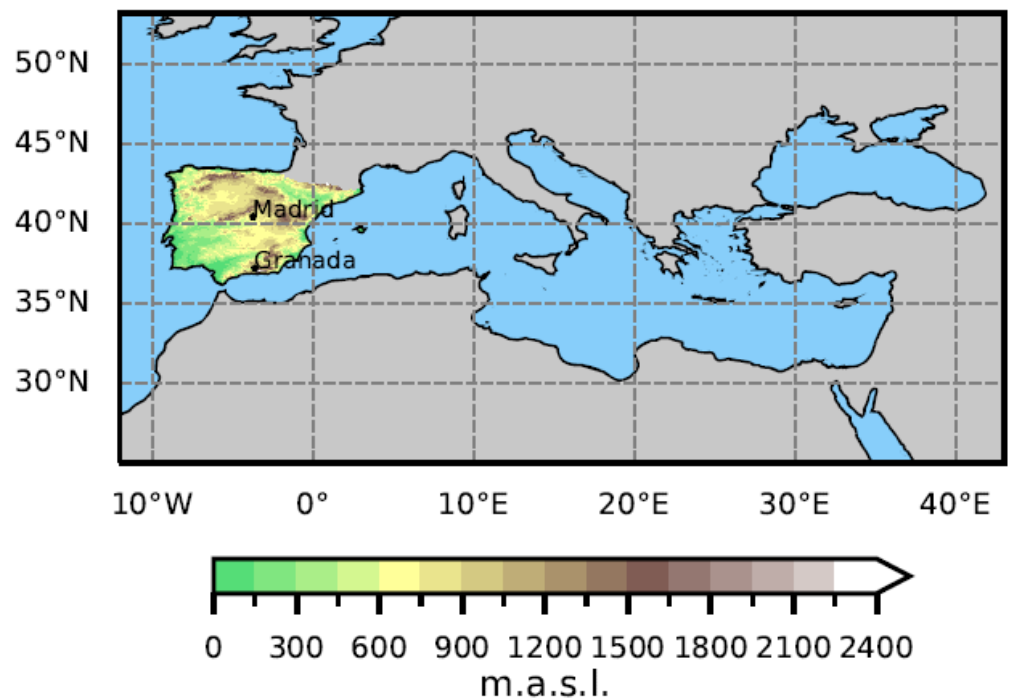

**Figure 1.** Location of the Iberian Peninsula within the Mediterranean sector. It includes the continental areas of Spain, Portugal, and Andorra.

The propagation from meteorological drought to agricultural drought is an understudied area, requiring further investigation to understand its complex characteristics. According to a recent review by [6], the co-occurrence of multiple driving factors in drought generation is a critical aspect that requires further attention. Currently, most studies only take into account soil moisture [7] or soil water deficit [8] and agricultural reservoir levels [9], and the combined contribution of soil moisture, evapotranspiration, and pre- cipitation to agricultural droughts is not yet fully understood. Accordingly, as suggested in previous studies, the development and application of new multi-variate indices [10] could be helpful in shedding light on the complex relationships between meteorological and agricultural drought. Additionally, for the analysis of propagation from meteorolog- ical to agricultural drought, it is appropriate to employ high resolution and long-term data, as highlighted by [11], in particular concerning soil moisture, which can be obtained through three major sources: in-situ observations, remote sensing, and hydrological models. The majority of studies based on in-situ observations involve measuring soil moisture levels at different depths across the globe using various soil moisture networks, but this approach has limitations as the observations are relatively short [12] and unevenly distributed [13], or may be unavailable in some isolated areas. Remote sensing products have been pre- ferred in some studies [14] since they provide better spatial coverage, but they appeared insufficient because they only cover a few centimeters of soil, while soil moisture from hydrological simulations has been found to be possibly affected by discrepancies compared to in-situ data [15]. Therefore, more studies that combine and evaluate different datasets are needed. Note that variations in vegetation health and/or cover may be due not only to rainfall or soil moisture deficits, but also to other stress factors, such as plant diseases. In this sense, indicators of vegetation stress and information on the deficit of precipitation and soil moisture must be considered together.

Given the existing research gaps, this study aimed to contribute to our understanding of agricultural drought over the IP, whose land cover is composed by a large extension of cropland along with other vegetation systems such as tree cover and grassland, specifically in relation to its propagation from meteorological droughts. To achieve this, a long-term dataset containing several drought events was analyzed, providing a comprehensive characterization of both meteorological and agricultural droughts. Various standardized drought indices were used, ranging from uni-variate to multi-variate indices that considered different physical quantities. In addition, a new combined index was proposed to account for the different factors that contribute to drought propagation. Overall, this study significantly advances our knowledge of meteorological and agricultural drought and their propagation process by leveraging a comprehensive set of tools, including meteorological, agricultural, multivariate, and combined drought indices. It offers a comprehensive perspective on the complex dynamics of drought, providing valuable insights for future research and informing effective drought management strategies.

## 2. Materials and Methods
### 2.1. Dataset

With the development of data assimilation technology, reanalysis data have become more representative of observed conditions and less limited than in-situ and remote sensing data. Reanalysis data offer global coverage, long time series, no gaps in space and time, and contain subsurface data, making them ideal for assessing agricultural drought. Several reanalysis datasets have been developed, and this study was conducted by employing the state-of-the-art reanalysis dataset for land applications, ERA5-Land [16], provided by the European Centre for Medium-Range Weather Forecasts (ECMWF) and included in the Copernicus Climate Change Service (C3S) of the European Commission. The ERA5-Land dataset was chosen as recommended in [17] due to its demonstrated relatively high accuracy compared to other remote sensing and reanalysis datasets [18] and hydrological models [19]. ERA5-Land offers a detailed record of hourly land surface evolution from several decades ago to the present, providing a vast array of key variables that represent the water and energy cycles. This dataset was chosen due to its superior ability to characterize the water cycle compared to ERA5 [20]. The original spatial resolution of the ERA5-Land dataset is 9 km on a reduced Gaussian grid, but C3S provides re-gridded data on a regular latitude-longitude grid of 0.1° × 0.1°, which corresponds to approximately 11 km at mid-latitudes. This study focused on the IP, covering a 72-year period from 1950 to 2021, using monthly-mean averages pre-calculated by C3S since sub-monthly fields were not necessary for our analysis.

To better explain the variables used in this study, we employed three key variables that are essential for computing drought indices:

- Total Precipitation [m]—This variable represents the total amount of rain and snow that has fallen on the Earth's surface between the beginning of the forecast time and the end of the forecast step. The units of precipitation are measured in depth in meters, which represents the extent of water that would be spread uniformly over the grid box;
- Soil Moisture [$m^3 m^{-3}$]—This variable represents the volume of water in different soil layers defined by the ECMWF Integrated Forecasting System. Specifically, for this study, we considered the first two ERA5-Land layers, which range from 0 to 28 cm in depth. Although the depth required for the most adequate representation of soil moisture content for agricultural droughts is still under exploration [21,22], we chose to follow the indication of [23], which suggests removing the lower layers to better represent the soil moisture conditions due to ancillary sources such as local rainfall or irrigation, so only the first two ERA5-Land layers were used (0–28 cm);

- Potential Evapotranspiration [m]—This variable is usually considered to be the amount of evaporation, under existing atmospheric conditions, from a surface of water having the temperature of the lowest layer of the atmosphere. The ECMWF Integrated Forecasting System computes it for an agricultural surface assuming it is well-irrigated, presuming that it does not significantly impact the atmospheric conditions in the region, such as humidity or cloud formation. This simplification allows for a standardized approach to estimating potential evaporation in agricultural contexts, which can introduce some uncertainties. In this respect, we compared the potential evapotranspiration derived from ERA5-Land with that calculated using the Penman-Monteith [24] equation, which takes into account various meteorological variables and thus eliminates the assumption of zero atmospheric impact, and we found that there were no significant differences in the results between the two methods.

*2.2. Drought Indices*

To identify meteorological droughts, we used the Standardized Precipitation and Evapotranspiration Index (SPEI) [10], which is based on the water balance of precipitation minus evapotranspiration and was chosen because it has been shown to be suitable for drought detection in Spain [3]. To capture agricultural droughts, we adopted several standardized indices to analyze the different outcomes generated by their distinct formulations. The first agricultural drought index was the Standardized Soil Moisture Index (SSI) [25], which was chosen for its simplicity and well documented capability to detect agricultural drought events [26], besides its reliability at a global scale for studying the propagation from meteorological drought detected by SPEI [13]. In addition, we adopted two indices to evaluate composite drought anomalies (agricultural and meteorological). We computed a multivariate index, the Multivariate Standardized Drought Index (MSDI) [27], which is based on the joint probability of precipitation and soil moisture, considering the effect of different variables in the characterization of agricultural droughts. We also included the Standard Precipitation, Evapotranspiration and Soil Moisture Index (SPESMI), a newly developed index that entirely accounts for the different variables involved in agricultural drought generation, introduced by [28] and formulated depending on both precipitation minus evapotranspiration balance and soil moisture. All indices used in this study were calculated using the non-parametric approach suggested by [25]. This method removes assumptions about the distribution of the variables and avoids the computationally expensive fitting of parametric distributions.

For example, MSDI was computed by treating precipitation and soil moisture at a selected time scale (e.g., 3 months) as random variables $X$ and $Y$, respectively, and considering their joint distribution,

$$P(X \leq x, Y \leq y) = p. \tag{1}$$

The empirical joint probability $p$ was estimated with the Gringorten plotting position formula [29] as in [25],

$$P(x_k, y_k) = \frac{m_k - 0.44}{n + 0.12}, \tag{2}$$

where $n$ was the number of the total input data and $m_k$ was the number of occurrences of the pair $(x_i, y_i)$ with $x_i \leq x_k$ and $y_i \leq y_k$ for $i\ .\ ,\ n$. Similarly, for univariate indices such as SSI the empirical marginal probability was calculated by using the univariate form of the Gringorten plotting position formula,

$$P(x_i) = \frac{i - 0.44}{n + 0.12},$$

where $n$ was again the number of total input data and $i$ was the rank of the observed values from the smallest. After obtaining the joint or marginal probability $p$ shown in Equation (1), to compute the drought index, it was only needed to retrieve the inverse of the standard normal distribution function $\varphi$ as in [25], namely,

$$MSDI = \varphi^{-1}(P). \tag{4}$$

By applying this methodology, we were able to calculate all the drought indices using the same approach, simply by modifying the variables used in the calculations. For example, in SPESMI, the joint probability *p* of precipitation minus evapotranspiration and soil moisture was calculated (see Table 1 for all the details).

**Table 1.** Characteristics of the standardized drought indices constructed with the non-parametric technique. P stands for Precipitation, SM for Soil Moisture, and E for Evapotranspiration.

| Drought Index | Structure    | Variables | Type of Drought     |
|---------------|--------------|-----------|---------------------|
| SPEI          | Multivariate | P–E       | Meteorological      |
| SSI           | Univariate   | SM        | Agricultural        |
| MSDI          | Multivariate | P, SM     | Agro-Meteorological |
| SPESMI        | Multivariate | P–E, SM   | Agro-Meteorological |

This not only made the calculations more straightforward but also ensured that the analysis was consistent across all the different indices. We used time scales of 1-, 3-, and 6-months to capture the immediate to seasonal/semi-annual impacts of precipitation, evapotranspiration, and soil moisture on drought characterization. To classify the drought categories, we followed the system described in [30] for the IP. Extreme drought was defined as indices with values below −2, severe drought as values between −2 and −1.5, and drought as values below −1 but above −1.5. Normal/wet conditions were associated with index values above 0, while dry conditions were indicated by an index value −0.5. See Table 2 for a summary of the categories.

**Table 2.** Drought categories for the uni-variate and the multi-variate standardized drought indices.

| Drought Index Value      | Drought Category | Conditions         |
|--------------------------|------------------|--------------------|
| Index $\leq$ −2          | −2               | Extreme Drought    |
| −2 < Index $\leq$ −1.5   | −1.5             | Severe Drought     |
| −1.5 < Index $\leq$ −1   | −1               | (Moderate) Drought |
| −1 < Index $\leq$ 0      | −0.5             | Dry                |
| Index > 0                | 1                | Normal/wet         |

Besides the described uni-variate and multi-variate indices, owing to the complex nature of drought events, another new index was proposed. In order to avoid relying on the information provided by a single index only, which might omit important characteristics of drought phenomena, a Combined Agricultural Drought Index (COMB) was developed adapting the Combined Drought Indicator described by [31]. In detail, COMB was based on the composition of the three agricultural drought indices SSI, MSDI, and SPESMI, and it was structured to favor the predominance of drought conditions over the other possible classes, namely when more than one indicator showed values below −1. To construct COMB, the drought categories reported for individual drought indices were taken into account, following the approach proposed by [32]. The index was not a simple average of the three indices, but rather a value was assigned to it based on the combination of drought categories. The severe drought condition (−1.5) was a refinement with respect to [32], to provide information with higher detail and consistency with [30]. When the three indices belonged to different categories, the arithmetic average was calculated and COMB was assigned to the resulting category. The focus of the combined index is on identifying drought conditions, while normal and wet conditions are only important for detecting the end of drought events. Please refer to Table 3 for details on the COMB classification.

**Table 3.** Methodology for the calculation of Combined Agricultural Drought Index (COMB).

| Drought Indices Values (SSI, MSDI, SPESMI) | COMB | Conditions |
| --- | --- | --- |
| 2 + indices $\in (-\infty, -2]$ | $-2$ | Extreme Drought |
| 2 + indices $\in (-2, -1.5]$ | $-1.5$ | Severe Drought |
| 2 + indices $\in (-1.5, -1]$ | $-1$ | (Moderate) Drought |
| 2 + indices $\in (-1, 0]$ | $-0.5$ | Dry |
| 2 + indices $\in (0, +\infty)$ | $1$ | Normal/wet |

### 2.3. Methods of Analysis

In this study, we investigated various aspects of drought phenomena, with a primary focus on the propagation from meteorological to agricultural drought. To this end, we employed different approaches to capture the distinct behaviors of the drought indices used. The first part of our analysis involved characterizing the two types of drought separately, with particular attention paid to agricultural drought events. We began by qualitatively examining the temporal evolution of the different drought indices at the three time scales, spanning the entire time window of the dataset (1950–2021). This involved observing the trends over the years and identifying possible similarities or differences between SPEI and the agricultural indices, as well as among the agricultural indices themselves.

In addition to characterizing meteorological and agricultural drought separately, we compared the agricultural drought indices to observations of vegetation health. To do so, we employed the Fraction of Absorbed Photosynthetically Active Radiation (FAPAR) Anomaly [33], which has been demonstrated to be effective in monitoring and assessing agricultural drought impacts [34]. Specifically, we used the FAPAR Anomaly indicator provided by the Copernicus European Drought Observatory EDO [35], which is computed as deviations from the long-term mean of biophysical FAPAR derived from surface reflectances measured by the MODIS-Terra satellite over a 21-year period (2001–2021). The EDO FAPAR anomalies are available at a spatial resolution of 1 km and are calculated for 10-day intervals. To compare these data to the monthly drought indices obtained from the ERA5-Land dataset, we calculated them on an 11 km grid and computed the mean for every 30-day period. Furthermore, following the recommendation of [36], a re-standardization step was performed on the monthly-averaged FAPAR anomalies obtained. This involved computing a new index, $F_s$, used to better compare and analyze the data and described by

$$F_s = \frac{F_i - \bar{F}}{\sigma}, \tag{5}$$

where for each specific year, $F_i$ represents the monthly averaged FAPAR anomaly for month $i$, while the mean and standard deviation of the monthly averaged FAPAR anomaly across all months $i$ during the entire time period from 2001 to 2021 are represented by $\bar{F}$ and $\sigma$, respectively.

The aim of the analysis using FAPAR anomalies was to assess the ability of agricultural indices to identify vegetation stress in areas affected by severe droughts. The analysis was based on verification metrics adapted from [37]. To compare the performance of the drought indices SSI, MSDI, SPESMI, and COMB with FAPAR anomalies, we used the following verification metrics: Probability Of Detection (*POD*), False Alarm Ratio (*FAR*), Critical Success Index (*CSI*), and Effect Of Drought (*EOD*). The drought categories presented in Tables 2 and 3 were considered, and the metrics were formulated as follows:

$$POD = H/(H + M)$$
$$FAR = F/(H + F)$$
$$CSI = H/(H + M + F)$$
$$EOD = (H + H_N)/(M + F + H + H_N),$$

where *H* (Hit) denoted the number of grids where the agricultural drought index showed categories 1, 1.5, or 2 and the FAPAR anomaly showed values belonging to the same range of categories; *M* (Miss) designated the number of grids where the FAPAR anomaly was subjected to categories −1, −1.5, or −2 and the agricultural drought index was subjected to categories higher than -1; $F$ (false alarm) stood for the number of grids where the FAPAR anomaly belonged to categories higher than −1, but the agricultural drought index indicated categories −1, −1.5, or −2; $H_N$ (Hit Null) expressed the amount of grids where the drought indices and FAPAR anomaly revealed categories 0.5 or 1. The total quantity of grids considered was given by the sum of $H$, $M$, $F$, and $H_N$, and the values of all four verification metrics ranged between 0 and 1, with a perfect fit characterized by $POD = 1$, $FAR = 0$, $CSI = 1$, and $EOD = 1$.

The subsequent step involved examining the relationship between the characteristics of droughts identified using different indices. To extract drought characteristics, we utilized the widely used *run theory* [38] to identify drought events. According to this method, a drought event begins when a drought index falls below a fixed threshold and continues until the index values remain continuously below that threshold (negative run), ending only when the index exceeds the threshold level (positive run). For this study, a threshold value of 1 was adopted for drought indices, as is customary. However, special attention was given to severe and extreme drought events, which are generally the most significant in terms of impacts and consequences.

After identifying the drought events, the next step involved calculating the average values of the agricultural drought indices on each gridpoint over the entire study period for the identified drought events. This was carried out to examine the relationship between the different types of indices and the properties of droughts. Specifically,

- Percentage: spatial fraction of IP affected by severe/extreme droughts;
- Duration: number of months with drought index values below the −1.5 threshold;
- Frequency: number of drought events per year;
- Severity: lowest value of the drought index during the drought periods;
- Intensity: ratio of drought severity to the drought duration;
- Magnitude: sum of absolute values of the drought index during the drought period.

After examining the characteristics of agricultural drought events using various indices, the second part of this study focused on analyzing the propagation of drought from meteorological to agricultural domains. To do this, the response time scale ($RT$) was calculated, following the approach of [39]. The $RT$ reflects the time it takes for the accumulated deficit in meteorological drought to correspond to agricultural drought, and it is based on a correlation analysis between agricultural and meteorological drought indices at different time scales. Specifically, for each location, the Pearson correlation was calculated between the time series of SSI, MSDI, SPESMI, and COMB at a 1-month time scale and SPEI at time scales of 1-, 2-, 3-, . . . , and 48-months. The response time scale for a given agricultural drought index was then determined as the time scale of SPEI with the highest correlation coefficient with the index values. This analysis covered the entire time period of the ERA5-Land dataset.

To analyze specific parameters of drought propagation, this study focused on a significant case study, the 2005 drought event in the IP, which had severe impacts on the entire European continent, reducing cereal yields by 10% [40]. This event is well documented and reported in various databases, including the European Drought Observatory and Emergency Events Database EM-DAT [41], making it a suitable reference for the analysis. The period including the 2005 drought was empirically investigated by examining the sequential spatial extent of drought coverage according to different drought indices at various time scales [42]. This approach allowed for the observation of spatial drought propagation across different systems over the IP, and some locations were selected for further analysis involving the lag time ($LT$) between the onset of meteorological and agricultural droughts. According to [43], the lag time for two different types of drought event was expressed as

$$LT = T_M - T_A, \qquad (6)$$

where $T_M$ and $T_A$ represented the initial time (in months) of meteorological and agricultural drought, respectively. Therefore, $LT$ was used to characterize drought propagation by measuring the difference in onset timing between the two drought episodes.

## 3. Results

### 3.1. Characterization of Agricultural Drought over the IP

The analysis began with a qualitative inspection of the indices computed over the time period from 1950 to 2021. Figure 2 displays the temporal evolution of the meteorological index (SPEI) and agricultural drought (AD) indices (MSDI, SPESMI, SSI, and COMB) at 1-, 3-, and 6-month time scales, which allowed for the distinction of the main features of meteorological and agricultural drought events. The evolution of the indices appeared smoother for larger time scales than for the 1-month time scale, and there was a slight trend towards increased dryness in the last two decades for both SPEI and AD indices, extending the assessment of droughts from meteorological to agricultural droughts. The analysis of Figure 2 revealed that the frequency and severity of meteorological drought episodes have increased since the 1990s, and the recurrence of severe droughts has notably developed in the last ten years. While the similarities of the indices suggested a good correlation between the contributions of precipitation deficit, water balance, and soil moisture, some differences emerged among the AD indices. SSI reached the highest values, while the two multi-variate indices reported the lowest values, with COMB being a compromise between the two types of indices. The multi-variate indices, which include the effect of precipitation, seemed to better reproduce the variations of SPEI, suggesting only a limited impact of soil moisture in their computation.

Figure 3 presents the skill score metric indices for the four AD indices at different time scales (1-, 3-, and 6-month) in comparison to FAPAR anomalies. The metrics include $H$ (hit) and $M$ (miss), which represent the ability of the AD indices to detect drought-affected areas where there were also dry vegetation conditions identified by FAPAR anomalies.
A high value of $POD = (H + M)/M > 0.8$ indicates that the AD indices were successful in identifying gridpoints dominated by agricultural droughts characterized by stressed land vegetation, especially for semi-annual droughts (6-month time scale) detected through MSDI and SPESMI. On the other hand, $F$ (false alarm) represents the number of grids where SSI, MSDI, SPESMI, or COMB detected drought conditions but with disagreement compared to the FAPAR anomalies. The non-zero results for $FAR = F/(H + F)$ indicate that there were some areas with dehydrated plants that could not be accurately monitored by AD indices, with almost negligible variations among the time scales and indices. The values of $CSI = H/(H + M + F)$ suggest that the regions classified with a desiccated flora by FAPAR anomalies were not only those where the AD indices identified drought conditions, while the opposite was, almost everywhere, true. The ratio of drought-affected areas detected by AD indices corresponding to high vegetation stress with respect to the total zones of high vegetation stress recognized by FAPAR anomalies was approximately 0.5, with slightly increasing CSI for semi-annual droughts. This suggests that FAPAR could monitor arid areas which could not be successfully captured by SSI, MSDI, SPESMI, or COMB, maybe due to the fact that FAPAR anomalies reveal variations in the vegetation health which can derive not only from rainfall or soil moisture deficits, but also from other stress factors such as plant diseases. Considering $H_N$ (hit null) as the amount of gridpoints that were free of droughts and with healthy foliage, the values of
$EOD = (H + H_N)/(M + F + H + H_N)$ lower than 1 implied some areas in mixed conditions, namely affected by drought, which had no relevant impacts on vegetation or unhealthy vegetated regions caused by factors other than droughts. In conclusion, SSI, MSDI, SPESMI, and COMB were efficient in assessing the vegetation stress of IP during drought events, but they were not sufficient to distinguish all the areas identified by FAPAR anomalies, whose stress could be due to different factors other than drought occurrence.

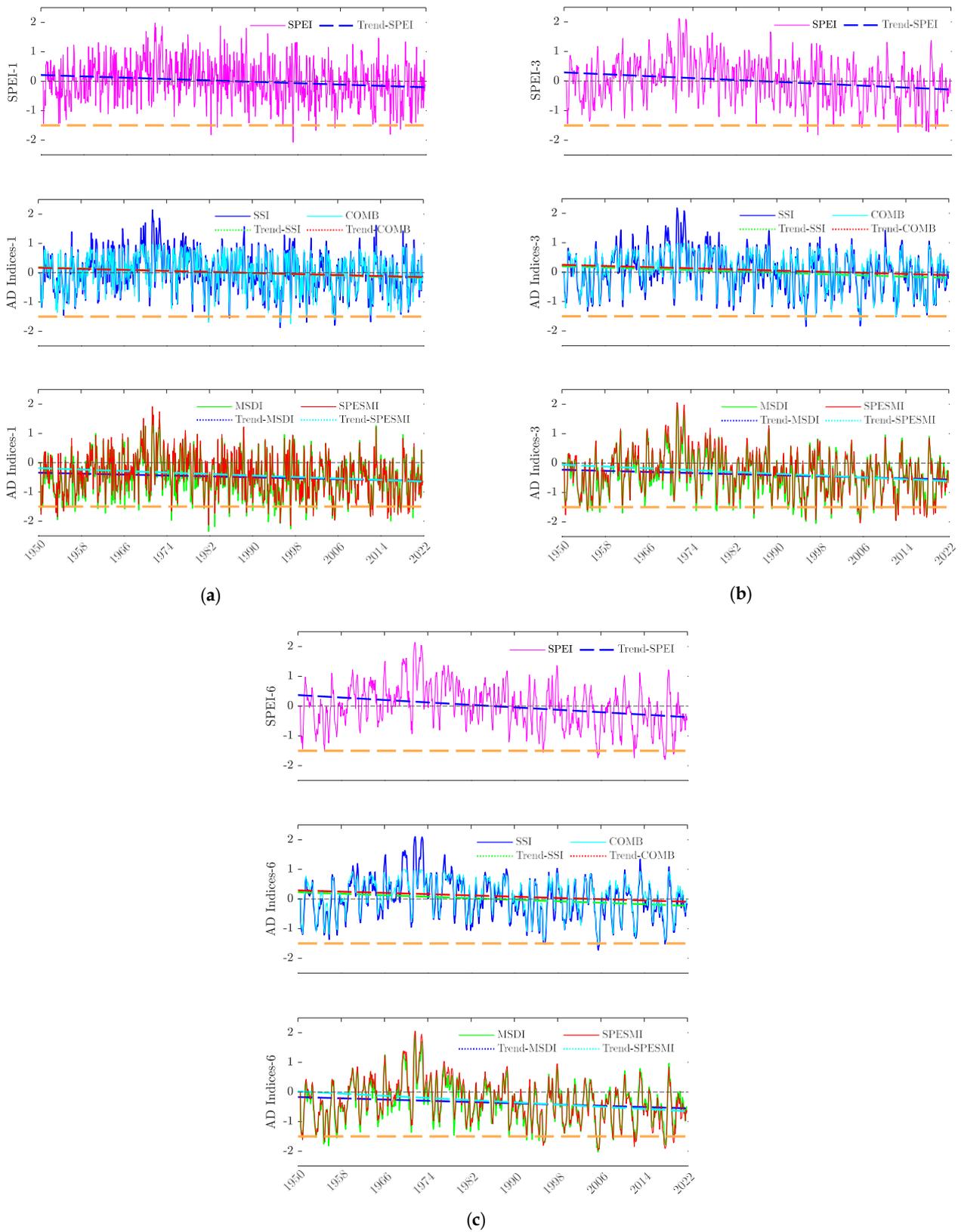

**Figure 2.** Temporal evolution of Iberian Peninsula averaged SPEI and agricultural drought (AD) indices (SSI, COMB, MSDI and SPESMI) at 1-month (**a**), 3-month (**b**) and 6-month (**c**) time scales for the entire period considered. The linear trend is also represented for each index. The orange dashed line indicates the −1.5 threshold for severe drought events.

To investigate the diverse responses of the AD indices in the IP region during the studied period, various drought characteristics were analyzed. Firstly, the average percentage of the IP region affected by drought events was computed. The results are illustrated in Figure 4, which displays the percentage of the IP region experiencing severe to extreme droughts (i.e., AD indices ≤ − 1.5) at 1-, 3-, and 6-month time scales. One key finding was the marked disparity between SSI and other indices, especially the two multi-variate indices. Specifically, while MSDI and SPESMI revealed more than 75% of the IP region suffering from droughts, SSI only covered less than 40%. The integration of precipitation (or water balance) and soil moisture deficits generated drought events in a wider area of the IP region compared to the case of soil moisture deficit alone, with a difference of approximately 40% for each time scale, a logical consequence of the MSDI's capability to detect both meteorological and agricultural events. COMB exhibited a percentage value of around 60%, a trade-off between uni-variate and multi-variate indices. The variations in the values of the same AD index depending on the time scale were small, and the highest percentage was generally observed for 1-month droughts.

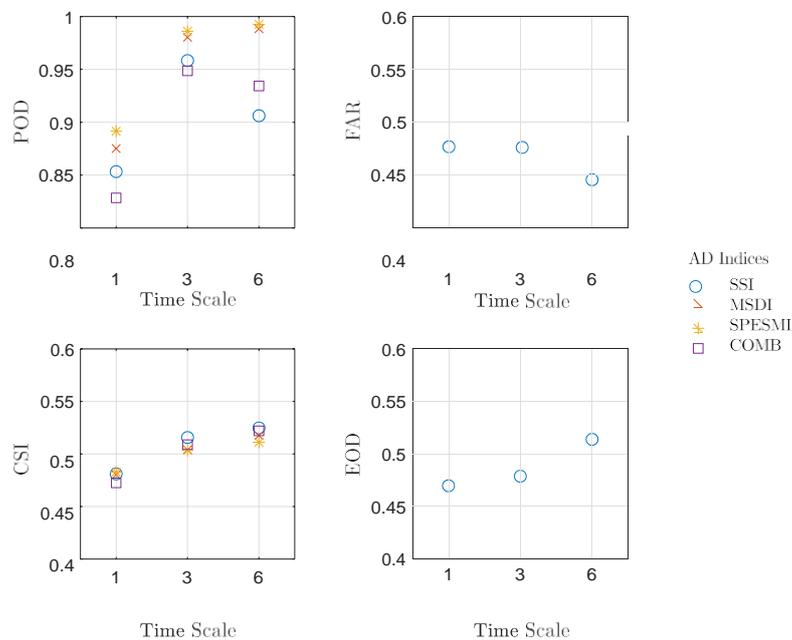

**Figure 3.** Skill score metrics regarding FAPAR anomalies for AD indices at the 1-, 3-, 6-month time scales.

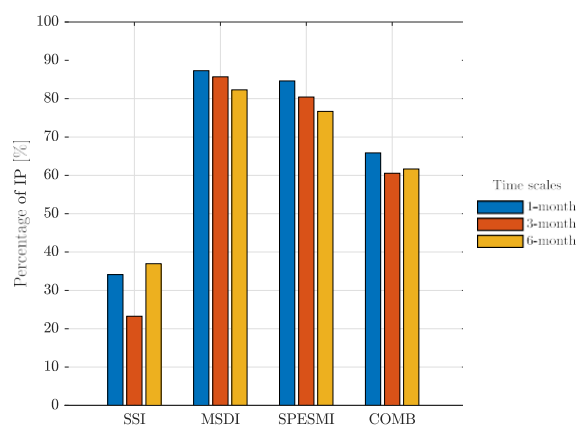

**Figure 4.** Percentage of the IP affected on average by severe/extreme droughts according to agricultural indices at the 1-, 3-, 6-month time scales.

To gain further insights into the agricultural droughts identified by the AD indices at different time scales, we investigated several other characteristics. One of these is the frequency of severe/extreme droughts per year, which is closely related to their duration, as well as their severity, which is typically used to assess the significance of drought events. Figure 5 displays these two characteristics for each index and time scale. The frequency of severe/extreme drought events per year (Figure 5a,c,e for 1-, 3- and 6-month time scales, respectively) confirmed that multi-variate indices were more sensitive than uni-variate ones in detecting drought events in a larger area. Specifically, SSI identified less than one drought event per year, on average, across all the IP at the 1-month time scale, while SPESMI and MSDI detected multiple drought episodes in several areas, with values ranging from around 1 event/year to 1.2 events/year on average. This is in agreement with the fact that precipitation, evapotranspiration, and soil moisture variables are vital factors to adequately represent agricultural drought conditions [28] and, especially in large areas with different climate characteristics, their combination by using multi-variate indices could better detect the occurrence of drought events. The patterns of COMB showed a balance between SSI and MSDI/SPESMI. The north-west and the Pyrenees were the most affected zones for all indices, indicating that both soil moisture and precipitation/water balance deficits were recurrent. Similar patterns were found at 3- and 6-month time scales, although the frequency values were naturally lower due to the longer duration of the considered events. The average drought severity showed a more complex behavior among the AD indices. SSI exhibited a large area of non-severe droughts (severity smaller than 1.5) at the 1-month time scale (Figure 5b), while COMB, SPESMI, and MSDI showed gradually increasing regions affected by severe drought conditions. However, the distribution of severity values differed among the indices. For instance, a small fraction of the centre-east of the IP was one of the most affected zones according to SSI but was not equivalently accounted for by the other indices, especially by MSDI, which retrieved its lowest severity values in the same area. This suggested a balance between the contribution of water balance variables in multi-variate indices, which could significantly impact the effect of soil moisture. Severity also showed a wider range of values for longer time scales. Bi-annual deficits in water balance and soil moisture resulted in large areas affected by severe droughts, with peaks close to extreme droughts in certain regions. On the other hand, even if the 6-month scale showed severity values similar to shorter time scales on average, severity appeared more pronounced or not depending on the region and the index. For example, in the south of Portugal, COMB-6, MSDI-6, and especially SPESMI described a higher severity than at 1- and 3-month time scales. Other areas, such as the mountain region in the southern part of the IP and the northern peninsular zone, reported the opposite behavior, revealing lower severity values at longer time scales.

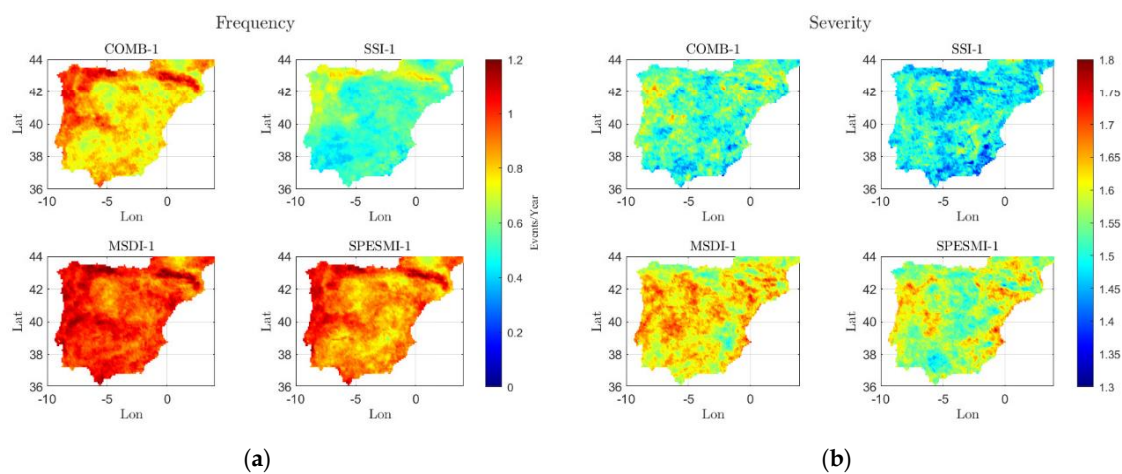

**Figure 5.** *Cont.*

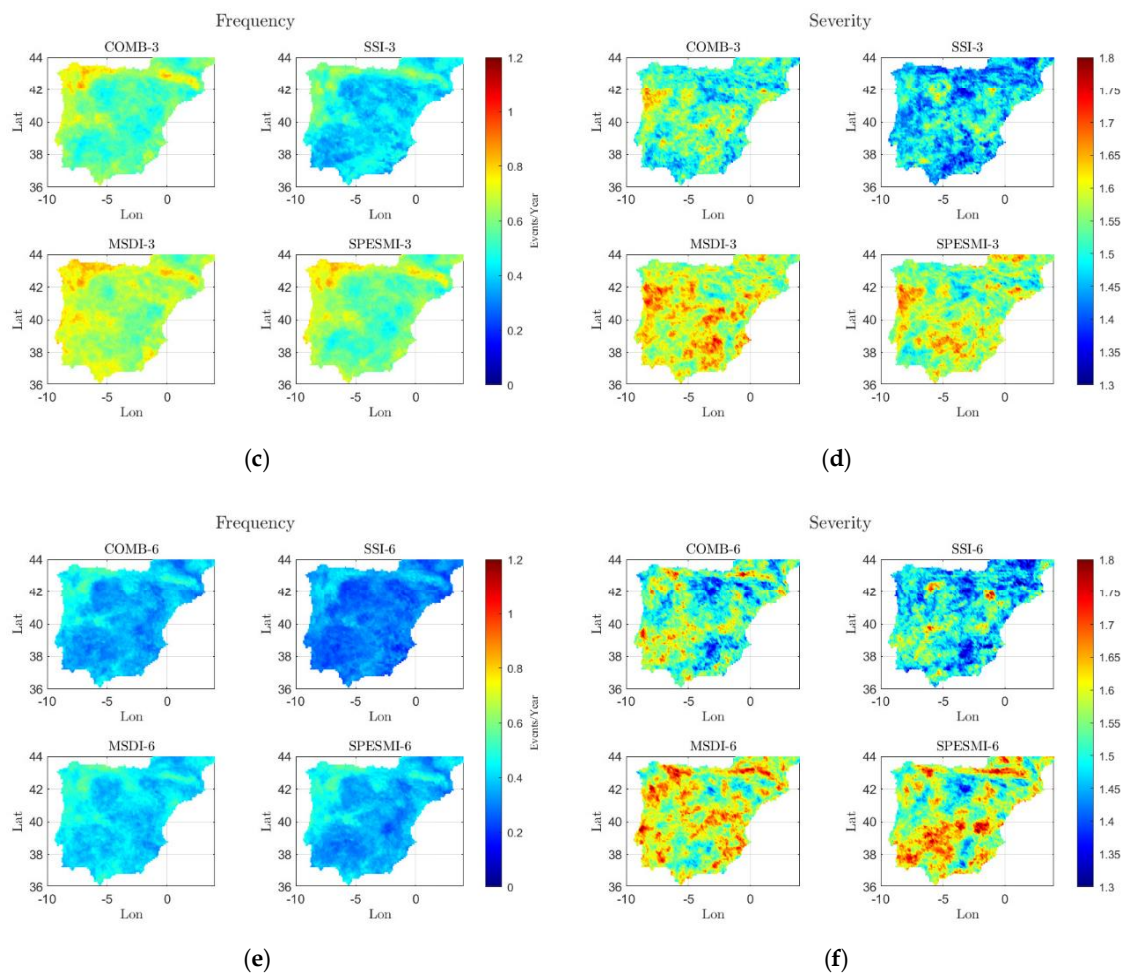

**Figure 5.** Patterns of the average characteristics of drought events on the Iberian Peninsula according to 1-, 3- and 6-month time scales for the different AD indices: frequency of severe/extreme droughts per year (**a**,**c**,**e**), and absolute value of drought severity (**b**,**d**,**f**).

*3.2. Propagation from Meteorological to Agricultural Drought*

This study investigated the response time scale (*RT*) as a key parameter in drought propagation, which represents the accumulated precipitation deficiency in the antecedent *RT* months that causes agricultural drought. A shorter *RT* indicates a faster response to meteorological drought. Figure 6a shows the maximum Pearson correlation between 1-month agricultural drought indices and SPEI computed from 1- to 48-month time scales for the entire 72-year period over the IP. Figure 6b indicates the corresponding *RT* in months for each gridpoint.

The analysis revealed a high correlation between agricultural and meteorological droughts at small time scales, consistent with previous studies [44]. The *RT* for SSI was 2 months for most of the IP, except for the Pyrenees where *RT* was 1 month and 3–4 months in some isolated regions, particularly in the southern coastal areas. MSDI, SPESMI, and COMB had an *RT* of 1 month, indicating that the contribution of other vari- ables accelerated the response compared to soil moisture alone. In particular, SPESMI, which includes evapotranspiration, presented the highest correlation with meteorological droughts detected by SPEI, which is also based on water balance.

For a detailed analysis of the propagation from meteorological to agricultural drought, we focused on the 2005 drought episode in the IP region. To investigate the spatial evolution of the drought phenomenon, we analyzed the monthly patterns of drought index values during the reference period from October 2004 to December 2005. Specifically, we examined the variations of the AD indices at the 1-month time scale and compared them to the progression of the SPEI at 1-month and 3-month time scales to understand the propagation from the current month and from the seasonal meteorological water balance deficit to the development of agricultural drought. Figure 7 illustrates the temporal evolution of the meteorological drought identified by the SPEI at 1- and 3-month time scales from October 2004 to December 2005 over the entire IP region.

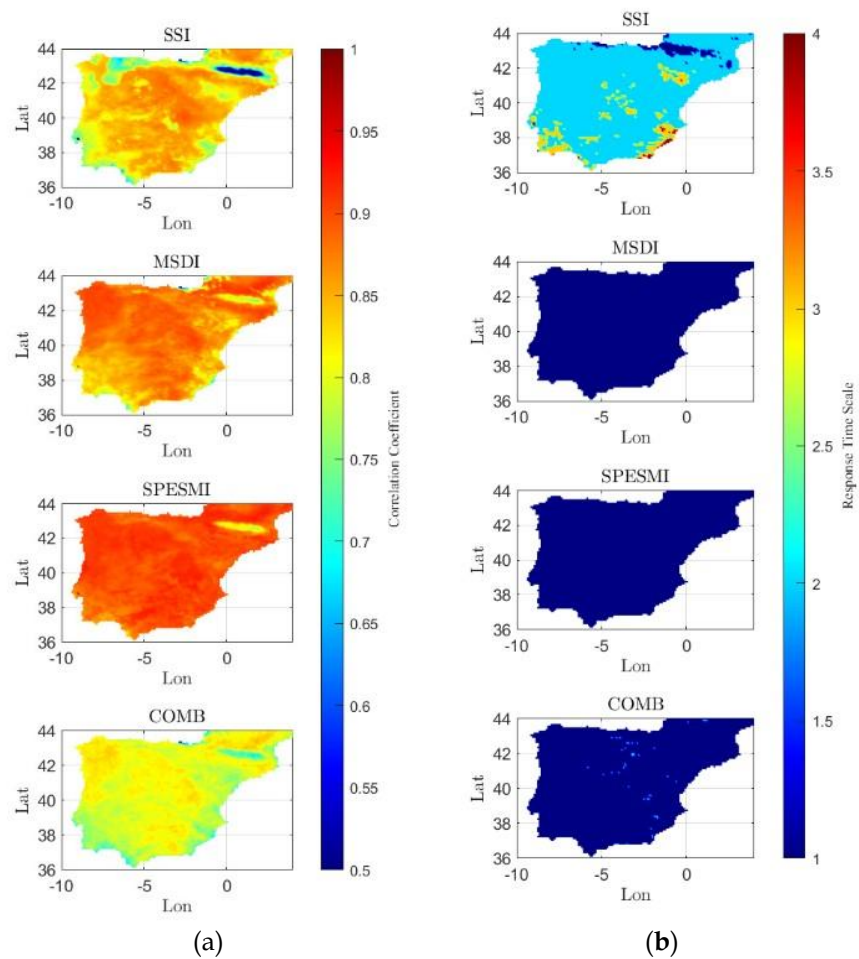

**Figure 6.** (**a**) Maximum Pearson correlation retrieved between the SPEI at the 1-, 2-,..., 48-month time scales and the 1-month time scale AD indices. (**b**) Corresponding *RT* in months for each gridpoint.

As shown in Figure 7a, the SPEI calculated at monthly intervals indicated a moderate drought condition in December 2004 over the majority of the IP. February 2005 was identified as the driest month, with a significant area experiencing extreme drought conditions. This pattern was followed by two wetter months, after which the severe/extreme drought episode reoccurred in May 2005, initially affecting only the south and eastern coasts of the IP. This condition lasted until the end of summer 2005 (September), with a modified pattern, which was more concentrated in the centre and northern IP. This event had severe impacts on Spain and Portugal, as reported by the European Drought Observatory, with a relaxation during the winter of 2005. The evolution of seasonal meteorological drought, shown in Figure 7b, was similar to the SPEI at monthly intervals, with the initial peak in February 2005. However, the SPEI-3 patterns were more continuous, essentially reporting prolonged drought conditions over the reference period, with a delayed conclusion (extreme drought conditions were observed even in October 2005). The driest region was the southern IP at the beginning, with varying features extending to the central and northern IP. To examine the progression of agricultural drought during the meteorological drought episodes, the spatial patterns of the four different agricultural drought indices at monthly time scale were analyzed. Figure 8 illustrates the spatial distribution of these indices across the IP during the reference period (October 2004–December 2005). All four AD indices captured the onset of agricultural drought in February 2005. However, there were some differences in their assessment of the drought conditions leading up to that point. While COMB, MSDI, and SPESMI indicated moderately dry conditions over the IP even from December 2004 except for the eastern coastal region, SSI showed normal or wet patterns during that period, suggesting that the incorporation of variables other than soil moisture may have allowed for a more accurate detection of drought impacts. Furthermore, SSI generally indicated less severe drought conditions than the other indices, and its patterns were less uniform and homogeneous compared to MSDI and SPESMI, particularly during the initial month of the drought event.

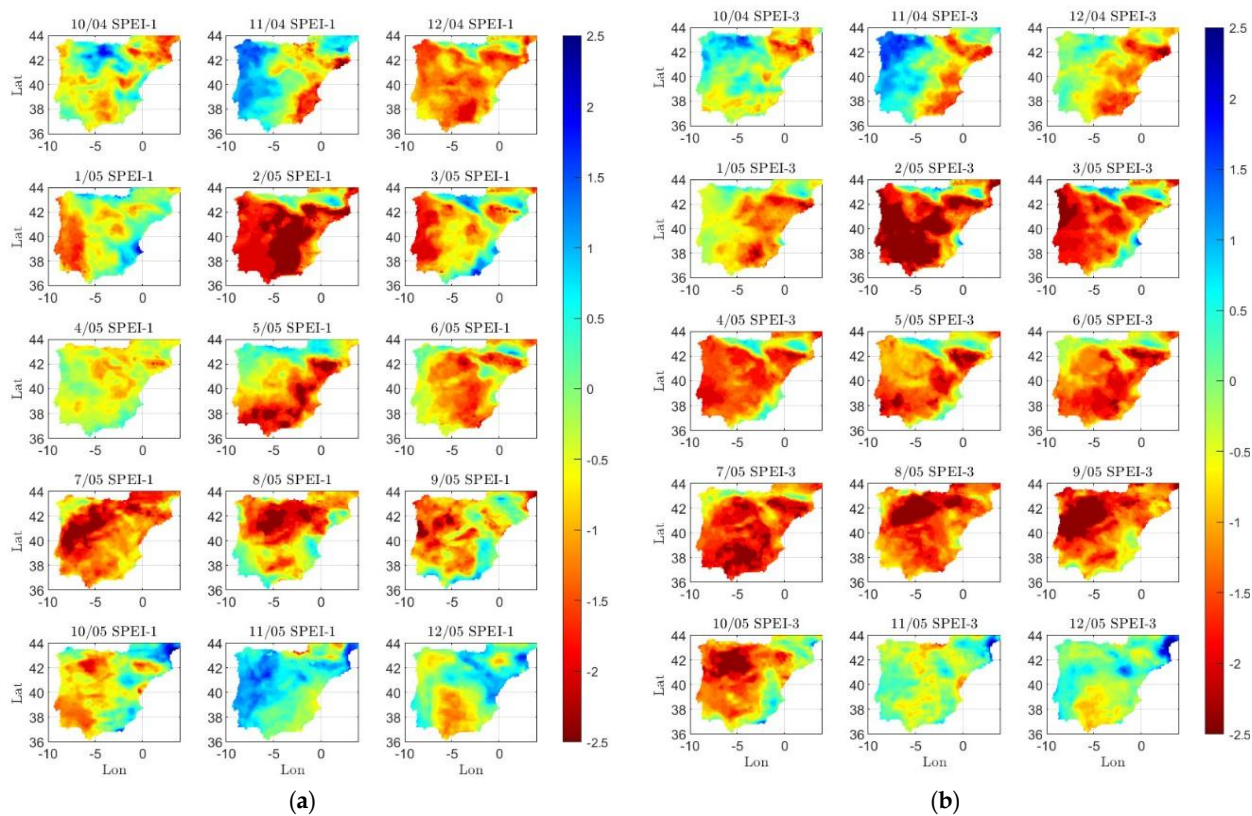

**Figure 7.** Temporal evolution of meteorological drought pattern over the IP according to SPEI at (**a**) 1- and (**b**) 3-month time scales, from October 2004 to December 2005.

MSDI and SPESMI showed a high degree of similarity in their behaviors, indicating that the inclusion of evapotranspiration had only a marginal impact on the results. COMB exhibited features that were balanced between the other three indices, but its similarity to the two multivariate indices appeared to have a greater influence on its performance than SSI. Overall, the patterns of agricultural drought identified by the AD indices were consistent with those identified by SPEI, particularly SPEI-3, which presented a higher spatial correlation with seasonal meteorological drought than with monthly water balance deficits. However, the severity values of the four AD indices were generally higher than those of SPEI-1 and SPEI-3, indicating that the agricultural drought impacts were more severe and extensive than the meteorological drought.

To understand the propagation of drought from meteorological to agricultural systems, Figures 7 and 8 were useful in providing a qualitative overview of the spatial details of the 2005 drought evolution. However, to derive more quantitative information, a further investigation was conducted. Drawing inspiration from [13], we analyzed the probability of drought propagation from meteorological to agricultural systems under different levels of severity. Specifically, the fraction of the IP experiencing agricultural drought conditioned on the occurrence of meteorological drought was calculated for each month between October 2004 and December 2005. This fraction was defined as the propagation probability (*PP*) and was computed separately for the four AD indices at the 1-month time scale, distinguishing between three severity thresholds, namely moderate, severe, and extreme drought. To provide a comprehensive understanding of the results, Figure 9 represents the temporal evolution of *PP* for each AD index. The panels in each row exhibit the *PP* values of agricultural droughts with gradually increasing severity from left to right, while the different colors refer to the severity threshold of meteorological drought, based on SPEI-1. For instance, the red line in the first panel of row one shows the time evolution of *PP* for the occurrence of moderate agricultural drought (based on SSI-1) conditioned on the occurrence of severe meteorological drought (based on SPEI-1).

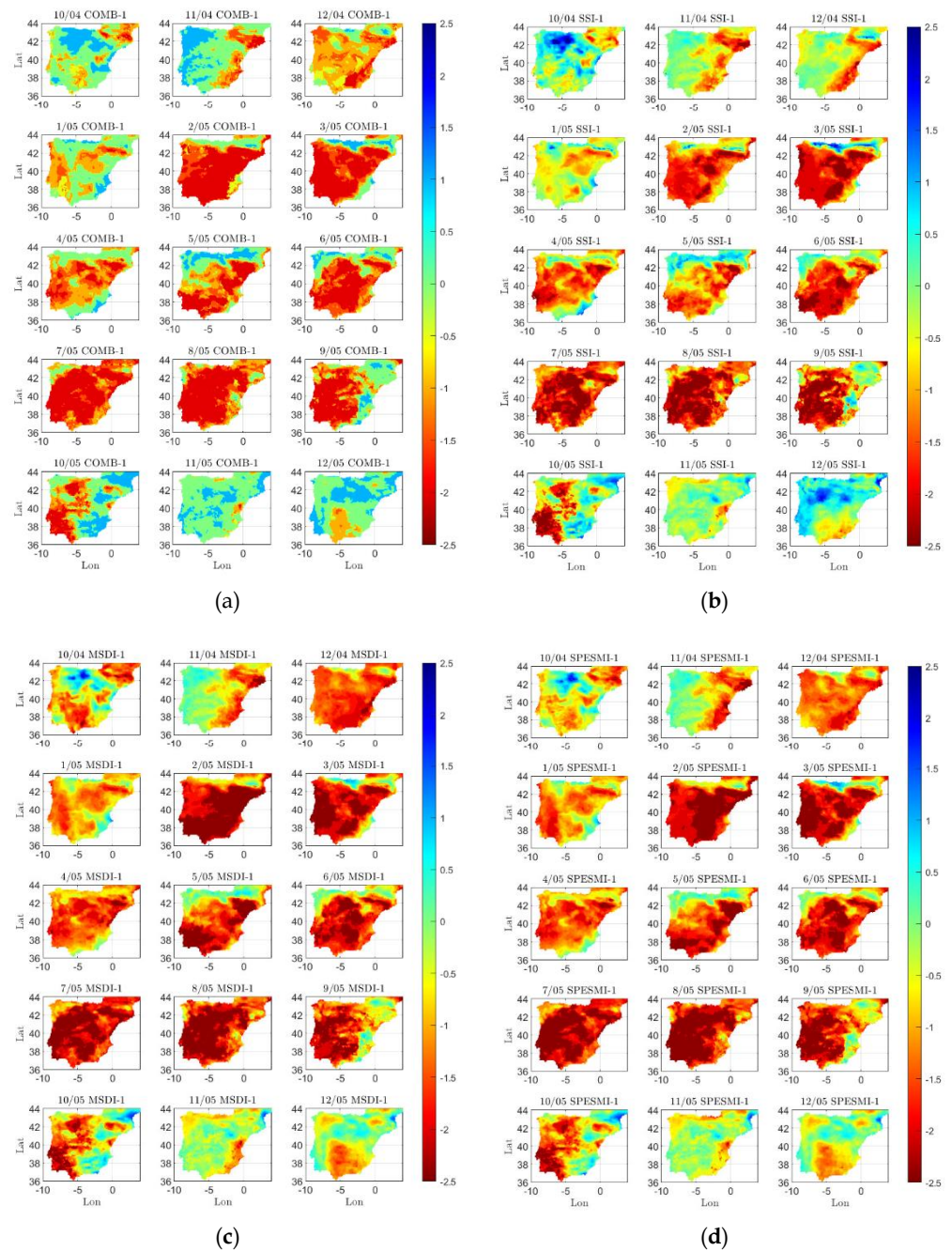

**Figure 8.** Temporal evolution of drought pattern over the IP according to the four AD indices: (**a**) COMB, (**b**) SSI, (**c**) MSDI, and (**d**) SPESMI at 1-month time scale from October 2004 to December 2005.

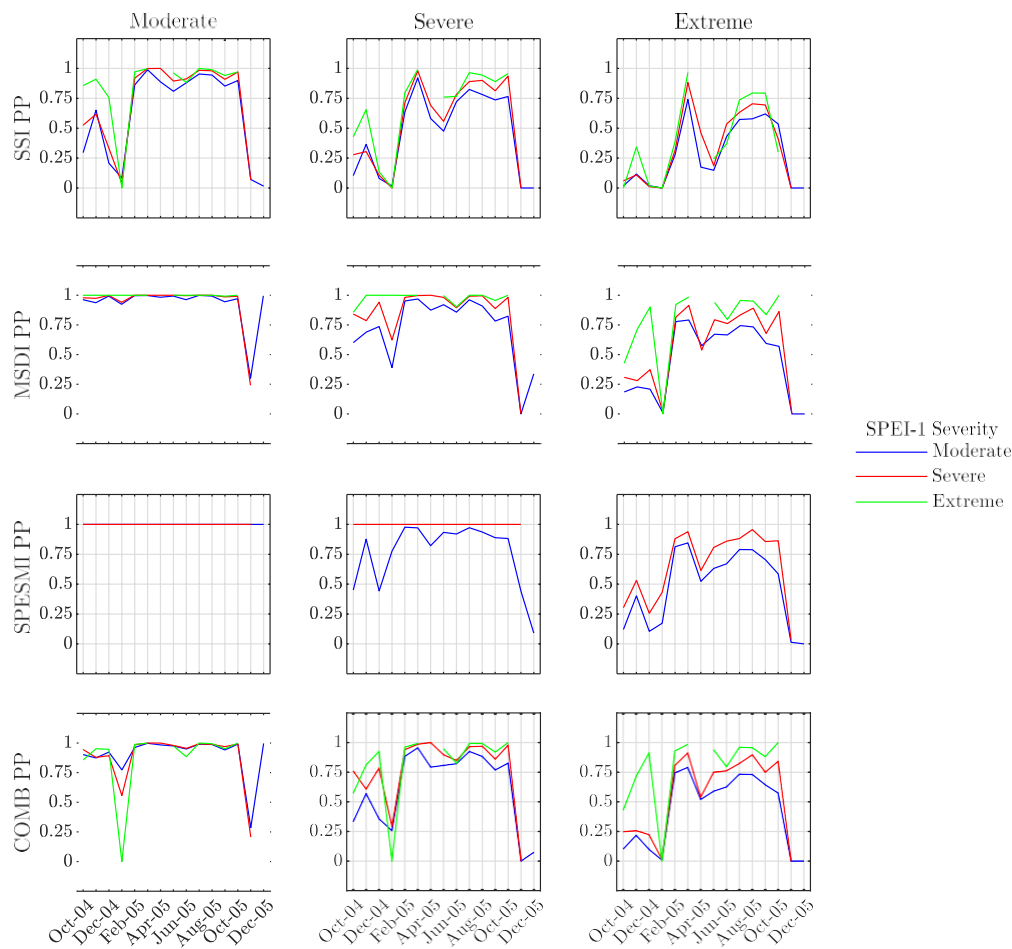

**Figure 9.** Propagation probability (*PP*) from meteorological drought detected with SPEI-1 to agricultural drought of different severity levels according to 1-month time scale AD indices. The colors distinguish the severity of generating meteorological drought (blue for SPEI-1 $\leq -1$, red for SPEI-1 $\leq -1.5$, green for SPEI-1 $\leq -2$).

The results of the analysis showed that all four AD indices displayed a significant increase in propagation probability (*PP*) as the severity levels of meteorological drought increased from moderate to extreme, with *PP* values approaching 1, indicating that the likelihood of agricultural drought was higher under drier meteorological conditions. For instance, in the first row of Figure 9, SSI indicated that areas affected by extreme meteorological drought from February 2005 were highly susceptible to various levels of agricultural drought, with the highest *PP* values observed for moderate severity propagation. MSDI and SPESMI consistently showed nearly constant *PP* = 1 values in the left panels, implying that these two multi-variate indices were highly sensitive to moderate agricultural drought propagation and less prone to severe and extreme propagation. On the other hand, SSI generally displayed lower *PP* values for all three severity levels compared to the multi-variate indices, while COMB demonstrated a balance between the two typologies of indices. Except for the deflection observed in March and April 2005, where the meteorological drought did not propagate in all the affected regions, all indices showed overall high *PP* values during the identified drought event, with *PP* > 0.5 for moderate SPEI-1 severity and *PP* > 0.75 for severe/extreme SPEI-1 severity. Although the *PP* values of MSDI, SPESMI, and COMB remained above 0.5 during this period, SSI demonstrated a more significant reduction, with *PP* < 0.25.

To investigate the propagation of drought from seasonal meteorological drought, we applied the same procedure as before. Figure 10 presents the behavior of $PP$ for agricultural drought conditioned on SPEI-3 values.

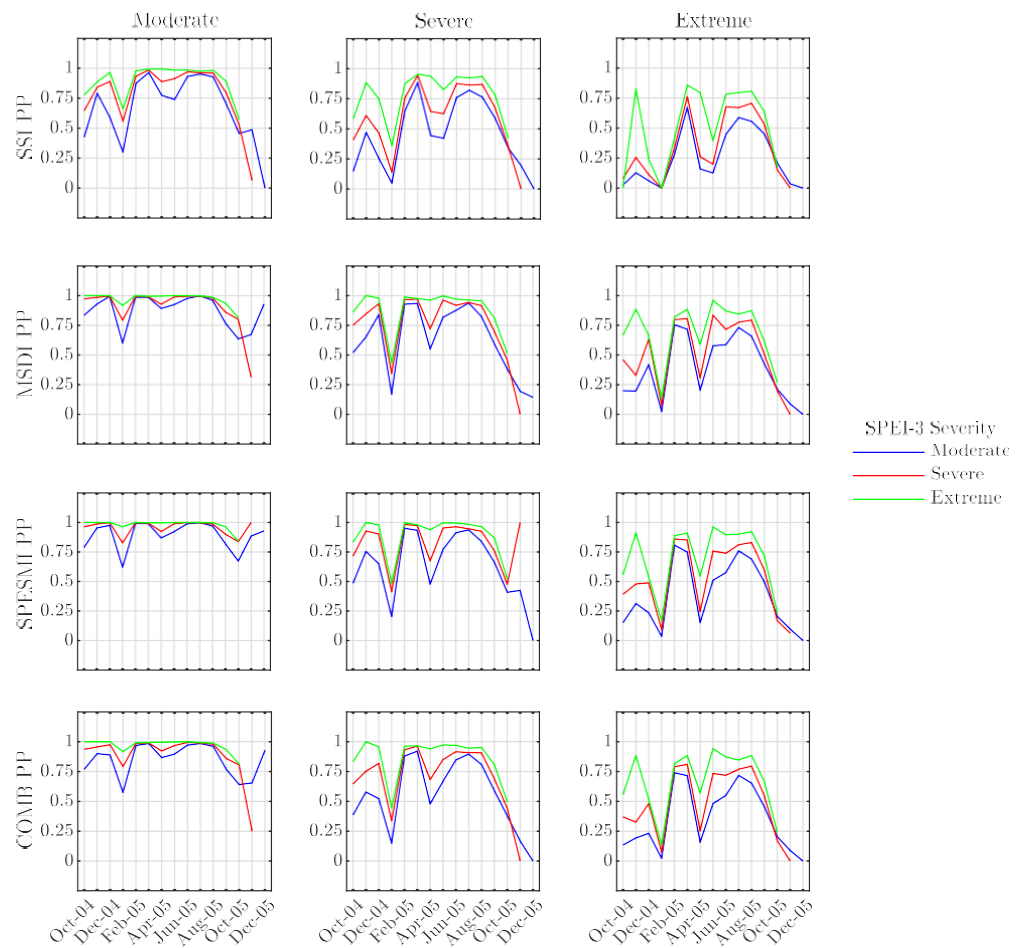

**Figure 10.** Propagation probability $PP$ from meteorological drought detected with SPEI-3 to agricultural drought of different severity levels according to 1-month time scale AD indices. The colors distinguish the severity of generating meteorological drought (blue for SPEI-3 $\leq -1$, red for SPEI-3 $\leq -1.5$, green for SPEI-3 $\leq -2$).

Similar to the analysis with monthly droughts, we observed a significant increase in $PP$ as the severity levels of meteorological drought enhanced from moderate to extreme. The maximum $PP$ values were found for moderate agricultural droughts, indicating that the propagation to more severe droughts was less likely to occur. Compared to the outcomes reported in Figure 9, the propagation from seasonal meteorological droughts revealed a more homogeneous behavior among the AD indices, with common features in the maxima/minima of $PP$ and the evolution of the $PP$ signal. Similar to the previous case, the lowest value of $PP$ was reached in March–April 2005, suggesting that, during this period, the meteorological drought was present but did not propagate into agricultural drought. Additionally, the panel regarding severe and extreme agricultural drought displayed smaller $PP$ values than the SPEI-1 example, indicating that seasonal drought was, in general, less efficient in propagation compared to drought caused by monthly water balance deficits.

To further investigate the 2005 drought event, two areas of the Iberian Peninsula that showed significant variations during the event were selected: the centre and the south. In order to monitor changes in drought indices and evaluate the lag time ($LT$), one city was chosen to represent each region. Following the approach of [45], Madrid was chosen as the representative location for the IP centre and Granada was selected for the IP South.

Figure 11 shows the 3-month time scale drought indices for these cities from January 2004 to January 2006. The seasonal drought was considered, as it is characterized by smoother and slower variations than the 1-month drought and provides more significant information for the calculation of *LT*.

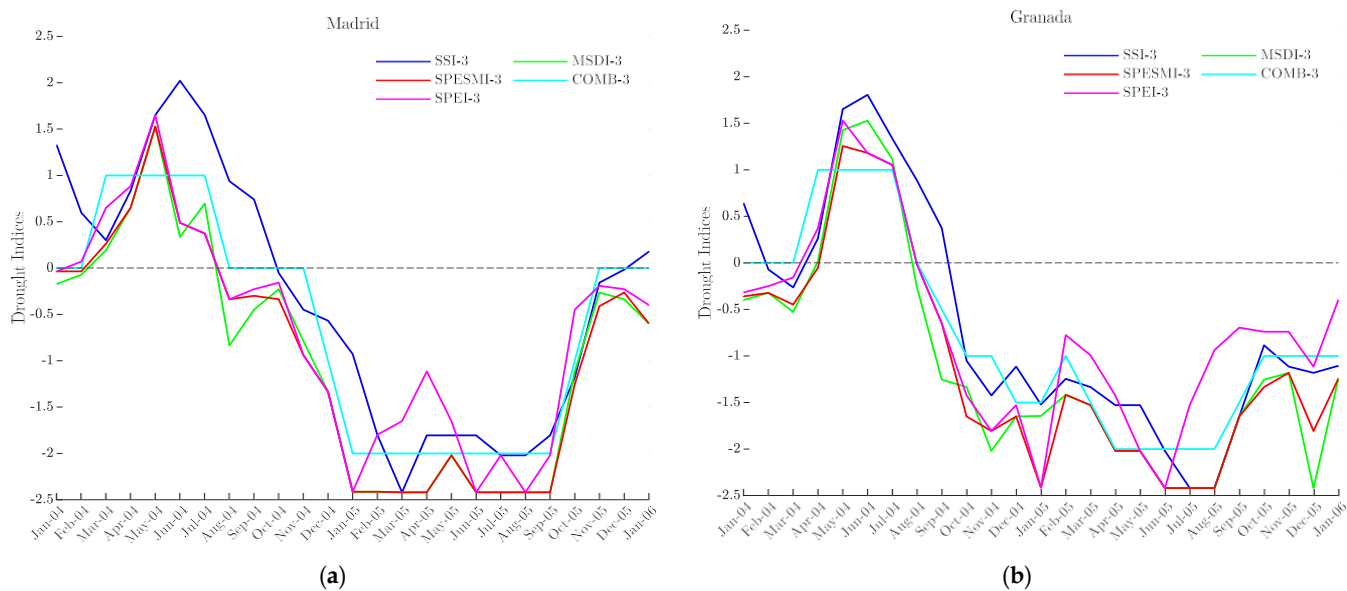

**Figure 11.** Local temporal evolution of different drought indices at 3-month time scale in Madrid (**a**) and Granada (**b**) from January 2004 to January 2006.

To calculate the *LT*, we compared the onset times of drought as measured by SPEI and the AD indices. In Madrid, the multivariate indices showed *LT* 0, indicating that they captured agricultural drought onset at the same time as SPEI. However, SSI had a lag time of about 2 months, since agricultural drought in that city began in February 2005, while meteorological drought began in December 2004. Compared to SPEI, the AD indices showed low variability, and changes in meteorological drought conditions did not necessarily result in modifications of agricultural droughts. The propagation of meteorological drought evolution had prolonged and almost constant effects on agricultural drought, especially according to COMB. The situation in Granada was slightly different. Meteorological drought onset preceded that in Madrid by 2 months (October 2004) and was simultaneous for all AD indices, resulting in *LT* 0. However, MSDI showed the onset of agricultural drought even 1 month before SPEI, indicating that the detected event was not solely associated with drought propagation. Although the onset lag time between meteorological and agricultural droughts was approximately zero, we noted that the duration of the two phenomena was not equivalent. While SPEI indicated dry meteorological conditions without the presence of drought in August 2005, SSI and COMB estimated the end of agricultural drought in October 2005, and MSDI and SPESMI required even more time.

## 4. Discussion

This study examined various types and temporal aspects of drought, with a specific focus on the transition from meteorological to agricultural drought, which marks the initial phase of drought propagation. In the first phase of the study, the aim was to characterize drought events across the IP over various time scales, including monthly, seasonal, and six-month periods, for the entire duration of the study period (1950–2021). The findings indicated a slight trend towards increased aridity during the last two decades. This was supported by both the SPEI, which aligns with the results obtained by [46] using parametric indices to identify meteorological droughts, as well as the AD indices. When comparing the values of the FAPAR anomaly, the AD indices were found to be effective in assessing vegetation stress during drought events in the IP. However, they were not always able to accurately detect all the dehydrated areas identified by FAPAR anomalies. In terms of the average characteristics of agricultural drought events, the SSI captured a noticeably smaller area affected by severe/extreme droughts (40%) compared to other AD indices (75%). The frequency of severe/extreme drought events per year also revealed that multi-variate AD indices were more sensitive than SSI, albeit with lower severity. Additionally, when considering long time scales, the severity of drought events showed a wider range of values than on the monthly time scale, ranging from the most to the least severe values. In the second phase of the study, we analyzed the mechanisms of propagation from meteorological to agricultural drought. The response time scale (*RT*) for the AD indices was calculated obtaining small values, consistent with findings from other studies [44]. Specifically, the results showed that the SSI had an *RT* of 2 months in most parts of the IP, while the MSDI, the SPESMI, and the COMB had an *RT* of 1 month, suggesting that the contribution of water balance accelerated the response compared to soil moisture alone. We analyzed the 2005 drought episode to study the temporal and spatial features of drought propagation in detail. The time evolution of seasonal agricultural drought, as revealed by the SSI, had a 2-month delayed onset compared to other AD indices (February 2005 vs. December 2004), while the end of the drought was almost coincident (October 2005). Furthermore, the severity values obtained by SSI were lower than those obtained by other AD indices. The agricultural drought patterns determined by the AD indices were consistent with those identified by SPEI, particularly showing a higher spatial correlation with seasonal rather than monthly meteorological drought. We evaluated the propagation probability (*PP*) from monthly meteorological drought and found that *PP* values were high during the identified drought event period. Monthly agricultural drought was also found to be more likely to occur when meteorological conditions were increasingly dry. However, smaller *PP* values were obtained from seasonal meteorological to monthly agricultural drought, indicating a reduced propagation efficiency compared to meteorological drought caused by monthly water balance deficits. The computation of lag time (*LT*) for SSI revealed different outcomes depending on the location, with values ranging from $LT \sim 0$ to $LT \sim 2$ months, while the multi-variate indices consistently showed $LT \sim 0$, regardless of the location. These values close to 0 suggested interest for a future analysis concerning sub-monthly features of *LT*.

There are currently several issues for analyzing drought propagation, such as the difficulty of comparing studies that use different indices and approaches, and the challenge of isolating single factors for analysis. Future studies on agricultural drought propagation should address these challenges by integrating techniques beyond statistical analysis based on run theory, such as extending the probabilistic approach of [47]. Considering the phenomenon of global warming, it is important to conduct studies that account for non-stationary conditions in future changing environments. In this regard, future work could expand upon the investigation by including the newly developed COMB index and incorporating an ensemble analysis with other meteorological and agricultural drought indices from existing literature. Additionally, future studies could explore the lead time between meteorological drought onset and agricultural drought onset, considering location and crop type. This would contribute to a more comprehensive understanding of agricultural drought propagation and enhance drought monitoring and early warning systems.

## 5. Conclusions

In conclusion, this study provided valuable information on the propagation of drought phenomena from meteorological to agricultural droughts on the IP. The investigation utilized a long record of data and distinct standardized indices, considering variables beyond just soil moisture such as precipitation and evapotranspiration. Some relevant outcomes were retrieved, in particular:

- Results showed a slight trend towards increased dryness over the last two decades and identified multi-variate AD indices as more effective in identifying severe drought events compared to the uni-variate SSI index.

- The severity values of the four AD indices were generally higher than those of SPEI-1 and SPEI-3, indicating that the agricultural drought impacts were more severe and extensive than the meteorological drought.
- This study introduced a novel combined agricultural drought index that balances the characteristics of other adopted indices and could be a valuable resource for future investigations.
- The response time scale was calculated for the AD indices and small values were obtained, also suggesting that the contribution of water balance accelerated the response compared to the effect of soil moisture alone.
- The analysis of the 2005 episode revealed a 2-month delayed onset compared to other AD indices in the analysis of seasonal agricultural drought, with a higher probability of propagation depending on the severity of the originating meteorological drought.

**Author Contributions:** Conceptualization, M.P., M.G.-V.O. and S.R.G.-F.; Data curation, M.P.; Formal analysis, M.P.; Funding acquisition, S.R.G.-F.; Investigation, M.P., M.G.-V.O. and S.R.G.-F.; Methodology, M.P. and M.G.-V.O.; Project administration, M.P. and M.G.-V.O.; Resources, M.P.; Software, M.P.; Supervision, M.G.-V.O. and S.R.G.-F.; Validation, M.P.; Visualization, M.P.; Writing—original draft, M.P.; Writing—review and editing, M.P. All authors have read and agreed to the published version of the manuscript.

**Funding:** This research has been carried out in the framework of the projects P20_00035 funded by FEDER/Junta de Andalucía-Consejería de Transformación Económica, Industria, Conocimiento y Universidades; LifeWatch-2019-10-UGR-01 co-funded by the Ministry of Science and Innovation through the FEDER funds from the Spanish Pluriregional Operational Program 2014–2020 (POPE) LifeWatch-ERIC action line; and PID2021-126401OB-I00, funded by MCIN/AEI/10.13039/501100011033/FEDER Una manera de hacer Europa.

**Data Availability Statement:** The data presented in this study are available on request from the first author.

**Acknowledgments:** The authors acknowledge Copernicus for providing open access data. The authors wish to express their appreciation to Silvana Di Sabatino for her instrumental role in initiating and facilitating a mutually beneficial collaboration between the Department of Physics and Astronomy at the University of Bologna and the Department of Applied Physics at the University of Granada.

**Conflicts of Interest:** The authors declare no conflict of interest.